# Superconductivity in a van der Waals layered quasicrystal

Yuki Tokumoto[1]*, Kotaro Hamano[1], Sunao Nakagawa[1], Yasushi Kamimura[1],

Shintaro Suzuki[2], Ryuji Tamura[3,4] & Keiichi Edagawa[1,4]*

[1]*Institute of Industrial Science, The University of Tokyo, Tokyo 153-8505, Japan*

[2] *Department of Physical Science, Aoyama Gakuin University, Kanagawa 252-5258,*

*Japan*

[3]*Department of Materials Science and Technology, Tokyo University of Science, Tokyo*

*125-8585, Japan*

[4]*CREST, Japan Science and Technology Agency (JST), Saitama 332-0012, Japan*

\* e-mail: tokumoto@iis.u-tokyo.ac.jp; edagawa@iis.u-tokyo.ac.jp

**Introductory paragraph**

van der Waals (vdW) layered transition-metal chalcogenides are attracting significant attention owing to their fascinating physical properties[1–5]. This group of materials consists of abundant members with various elements, having a variety of different structures. However, all vdW layered materials studied to date have been limited to crystalline materials, and the physical properties of vdW layered quasicrystals have not yet been reported. Here, we report on the discovery of superconductivity in a vdW layered quasicrystal of $Ta_{1.6}Te$ (refs. 6,7). The electrical resistivity, magnetic susceptibility, and specific heat of the $Ta_{1.6}Te$ quasicrystal fabricated by reaction sintering, unambiguously validated the occurrence of bulk superconductivity at a transition temperature of ~1 K. This discovery can pioneer new research on assessing the physical properties of vdW layered quasicrystals as well as two-dimensional quasicrystals; moreover, it paves the way toward new frontiers of superconductivity in thermodynamically stable quasicrystals, which has been the predominant challenge facing condensed matter physics since the discovery of quasicrystals almost four decades ago.

In 1984, Shechtman et al. discovered an exotic solid phase in a rapidly quenched Al–Mn alloy, which exhibited an electron diffraction pattern comprising sharp Bragg spots arranged in ten-fold rotational symmetry[8]. The sharp diffraction spots indicated the existence of a long-range translational order, and the ten-fold symmetric diffraction pattern was incompatible with a periodic order, thus verifying that the phase could not be crystalline. Soon after this discovery, a new classification scheme for solids was proposed, with the aforementioned phase denoted a quasicrystal (QC)[9–11]. Essentially, QCs have a quasiperiodic long-range translational order that is compatible with crystallographically disallowed rotational symmetries such as five-, ten-, and twelve-fold symmetries. Extensive studies have subsequently revealed that QCs are not exceptional but rather ubiquitous, as they have been formed in several binary and ternary systems. Of the 70 QCs that have been reported to date, approximately 40 have been found to be thermodynamically stable[12].

In 1998, Conrad et al. discovered a unique stable QC phase exhibiting dodecagonal symmetry in the compound of $Ta_{1.6}Te$ (ref. 6). This new QC has several notable features that have not been observed to date. First, this phase is the only reported transition-metal chalcogenide QC. Second, this is the only QC of a vdW layered material; its structure is realized through periodic stacking of two-dimensional dodecagonal (dd)-QCs induced by vdW interactions. The structures of this dd-QC phase and the related crystal approximant (CA) phases have been examined by X-ray diffractometry (XRD), electron diffractometry, and high-resolution transmission electron microscopy[6,7,13–15]. Moreover, a few layers of the dd-QC phase have recently been isolated using standard exfoliation techniques[7]. However, the number of studies on this phase is still very limited because arc-melting, which is usually used for fabrication of QC alloys, cannot be applied since the melting temperature of Ta is much higher than the boiling temperature of Te; to the best of our knowledge, experimental studies on elucidating the physical properties of the $Ta_{1.6}Te$ dd-QC phase have not been reported.

Therefore, in this study, the electrical resistivity, magnetic susceptibility, and specific heat of a polygrain single-phase $Ta_{1.6}Te$ dd-QC were measured down to low temperatures. The results categorically confirmed the occurrence of bulk superconductivity in a thermodynamically stable QC for the first time at a transition temperature $T_c$ of ~1 K. Superconductivity has never been reported for QC phases, except for a recent example of a metastable icosahedral QC (i-QC) phase of Al–Zn–Mg, which demonstrated bulk superconductivity at $T_c \approx 0.05$ K (ref. 16). Recent theoretical studies have suggested that the quasicrystalline superconductivity exhibits unconventional behavior such as the nonzero sum of the momenta of Cooper pair electrons[17] and intrinsic vortex pinning not by impurities and defects[18]. Furthermore, the specific heat and *I–V* characteristic curves of a normal-metal–superconductor tunnel junction have also been expected to display unusual tendencies[19].

Polygrain single-phase samples of the $Ta_{1.6}Te$ dd-QC were fabricated by reaction sintering, powdered, and then subjected to electron diffractometry and XRD analyses (see Methods section). The electron diffraction pattern of the synthesized sample (Fig. 1a) shows an arrangement of sharp spots with dodecagonal symmetry, which is essentially identical to that previously reported for the $Ta_{1.6}Te$ dd-QC phase[6]. A powder XRD profile of the sample was acquired using Cu Kα radiation (Fig. 1b); the Cu K$\alpha_2$ component was removed numerically, yielding a profile that corresponded to the Cu K$\alpha_1$ component (wavelength: 1.5405 Å). A comparison of this profile with those of the crystalline phases of $TaTe_2$ and Ta—the materials used for the reaction sintering (Supplementary Fig. 1)— revealed no traces of these phases in the acquired profile, indicating the completion of the reaction and the formation of another phase. However, because the powder XRD profile of the $Ta_{1.6}Te$ dd-QC phase has not been previously reported, it was constructed based on that of the CA phase of $Ta_{97}Te_{60}$, whose structure has previously been determined through single-crystal XRD measurements[14]. In general, by introducing phason strain to a QC structure, we can obtain a CA structure. Using this relationship, we can calculate the powder XRD profile of a QC phase from that of its CA phase. The principles and procedures underlying the calculation are detailed in Supplementary Section 1.2. The acquired and constructed profiles were compared (Fig. 1b and the magnified profiles in Fig. 1c). Although the peaks in the acquired profile were not adequately resolved in the $2\theta$ region of 33°–47°, the overall agreement was satisfactory; this indicated that the sample comprised entirely of the $Ta_{1.6}Te$ dd-QC phase. We note that comparing the acquired XRD profile with those of the $Ta_{21}Te_{13}$ and $Ta_{97}Te_{60}$ CA phases (Supplementary Fig. 9) showed clear discrepancies. The peak indices shown in Fig. 1c refer to the five basis vectors $\mathbf{a}_i^* = a^*(\cos\left(\frac{2\pi(i-1)}{12}\right), \sin\left(\frac{2\pi(i-1)}{12}\right), 0)$ ($i$ = 1–4) and $\mathbf{c}^* = (0, 0, c^*)$, with $a^*$ and $c^*$ being 0.6942 Å$^{-1}$ and 0.6047 Å$^{-1}$, respectively (Supplementary Table 1). Complete data on the powder XRD profile of the $Ta_{1.6}Te$ dd-QC phase are provided in Supplementary Table 3.

The temperature dependence of the electrical resistivity normalized by the value at $T = 300$ K—$\rho/\rho_{300\ K}$—was analyzed for three different samples (#1–#3; Fig. 2a); the three samples were prepared in the same way and used for repeatability testing. $\rho_{300\ K}$ was approximately 1.7 mΩ·cm for all the samples. $\rho(T)$ gradually increased with decreasing temperature and abruptly descended to zero at a low temperature, indicating the emergence of superconductivity. Analysis of the temperature dependence of $\rho/\rho_{300\,K}$ below 2 K (Fig. 2b) indicated that $\rho(T)$ abruptly dropped to zero at a midpoint transition temperature $T_c^{\mathrm{mid}} = 0.98$ K for all three samples, where $\rho(T_c^{\mathrm{mid}}) = 0.5 \times \rho(1.1\,\mathrm{K})$. Examination of the magnetic field dependence of $\rho/\rho_{300\,K}$ for Sample #1 at different temperatures (Fig. 2c) suggests that the superconducting transitions were suppressed by increasing the magnetic field. Similar results were obtained for Samples #2 and #3. By applying the criterion of 50% of the normal state resistivity to define the upper critical field $H_{c2}$, the temperature dependence of $H_{c2}$ was explored (Fig. 2d) by plotting $-H_{c2}/[T_c(dH_{c2}/dT)_{T=T_c}]$ against $T/T_c$ (that is, $h^*$ vs. $t$). According to the Werthamer–Helfand–Hohenberg (WHH) theory[20], which considers spin–orbit scattering and spin paramagnetism, $H_{c2}$ in the dirty limit (mean free path $l \ll$ coherence length $\xi_0$) is expressed as follows, in terms of the digamma function:

$$\ln\frac{1}{t} = \left(\frac{1}{2} + \frac{i\lambda_{SO}}{4\gamma}\right)\psi\left(\frac{1}{2} + \frac{\bar{h} + \frac{1}{2}\lambda_{SO} + i\gamma}{2t}\right) + \left(\frac{1}{2} - \frac{i\lambda_{SO}}{4\gamma}\right)\psi\left(\frac{1}{2} + \frac{\bar{h} + \frac{1}{2}\lambda_{SO} - i\gamma}{2t}\right) - \psi\left(\frac{1}{2}\right), (1)$$

where $\psi$ is the digamma function, $\gamma \equiv \left[(\alpha\bar{h})^2 - \left(\frac{1}{2}\lambda_{SO}\right)^2\right]^{\frac{1}{2}}$, and $\bar{h} = \frac{4}{\pi^2}h^* = \frac{4H_{c2}}{\pi^2(-dH_{c2}/dt)_{t=1}}$, with the parameters $\lambda_{SO}$ and $\alpha$ representing the effects of the spin–orbit scattering and spin paramagnetism, respectively. The obtained data were adequately fit to equation (1) using $\alpha = \lambda_{SO} = 0$, indicating that the two effects were negligible. Through this fitting, the value of the zero-temperature upper critical field $H_{c2}(0)$ was deduced to be 32 kOe. The dirty-limit superconductivity was consistent with the large residual resistivity shown in Fig. 2a. A similar $H_{c2}$ tendency has been observed for the Al–Zn–Mg i-QC[16].

The magnetic susceptibility and specific heat were investigated to corroborate the occurrence of the bulk superconductivity. The temperature dependence of the magnetic susceptibility $\chi$ was monitored at an external magnetic field of 50 Oe under zero-field cooling (ZFC) and field cooling (FC) conditions (Fig. 3a). In the ZFC curve, a sharp drop occurred at an onset temperature of 0.95 K—a value consistent with the transition temperature corresponding to $\rho(T)$ in Fig. 2b; moreover, a large diamagnetism was observed, indicating the exclusion of the magnetic flux owing to the Meissner effect. The low-temperature data in Fig. 3a suggest the existence of the relationship $4\pi\chi \approx -1$ emu/(Oe·cm$^3$), which corresponds to a superconducting volume fraction of ~100%, thereby providing solid evidence of bulk superconductivity. The XRD data (Figs. 1b and 1c) confirmed that this bulk superconductivity emanated from the dd-QC phase.

The temperature dependence of the specific heat $C$ was probed by plotting $C/T$ against $T^2$ (Fig. 3b). A jump in the $C$ value was clearly observed at $T \approx 1$ K—a value consistent with the transition temperatures corresponding to $\rho(T)$ (Fig. 2b) and $\chi(T)$ (Fig. 3a). The linear behavior of $C/T$ versus $T^2$—that is, the relationship $C/T = \gamma + \beta T^2$—was observed at $T^2$ values above ~1.4 K$^2$; accordingly, the corresponding $\gamma$ and $\beta$ values were estimated as 0.0224 mJ/(g·K$^2$) and 0.00532 mJ/(g·K$^4$), respectively. The electronic part of the specific heat $C_{el}$ was determined by subtracting the lattice contribution of the specific heat $\beta T^3$ from the measured specific heat, and $C_{el}/\gamma T$ was subsequently plotted against $T$ (Fig. 3c). Although the jump height $\Delta C_{el}$ at $T_c$ could not be precisely evaluated owing to the lower limit of the measurement temperature, the expression $\Delta C_{el} \gtrsim 1.0\,\gamma T_c$ was deemed to apply, which is close to the jump height $\Delta C_{el} = 1.43\,\gamma T_c$ expected from the Bardeen–Cooper–Schrieffer (BCS) theory. This large jump further verified the occurrence of the bulk superconductivity.

Using the aforementioned value of $\beta$, the Debye temperature $\Theta_D$ was deduced to be 132 K. Furthermore, the electron–phonon coupling constant $\lambda_{ep}$ was obtained using the McMillan equation[21]:

$$T_c = \frac{\Theta_D}{1.45} \exp\left\{-\frac{1.04(1 + \lambda_{ep})}{\lambda_{ep} - \mu^*(1 + 0.62\lambda_{ep})}\right\}, \qquad (2)$$

where $\mu^*$ denotes the Coulomb pseudopotential parameter, which is typically assumed to be 0.13 for superconductors including transition metals. Using $T_c = 0.98$ K and $\Theta_D = 132$ K, $\lambda_{ep}$ was calculated to be 0.52, indicating a weak-coupling superconductivity. In general, the density of a QC phase is approximately the same as that of its CA phase. Therefore, the density of the $Ta_{1.6}Te$ dd-QC phase can be estimated to be 10.6 g/cm³, calculated for the $Ta_{97}Te_{60}$ CA phase from its structure data[14]. Using this value and the estimated values of $\gamma$ and $\lambda_{ep}$, the density of states (for one of the two spin states) at the Fermi level $D(E_F)$ was estimated as $1.24 \times 10^{47}$ states/(J·m³) using the expression[21]

$$D(E_F) = \frac{3\gamma}{2\pi^2 k_B^2 (1 + \lambda_{ep})}, \qquad (3)$$

where $k_B$ is the Boltzmann constant.

Crystalline Ta–chalcogenide systems including $Ta_2Se$, $TaSe_2$, and $TaS_2$ have been found to exhibit superconductivity at ambient pressure[22,23]; however, the Ta–Te system has not been reported in this regard. The superconducting parameters of several crystalline Ta–chalcogenide systems[22,23], the $Ta_{1.6}Te$ dd-QC, Ta (ref. 24), and Al–Zn–Mg i-QC[16] are summarized in Table 1. For a simple metal system of the Al–Zn–Mg i-QC, $\mu^* = 0.1$ was used[21] in the calculation of $\lambda_{ep}$, while for the others $\mu^* = 0.13$ was used. The $\lambda_{ep}$ values of Ta and $Ta_2Se$ were fairly high, which may be in an intermediate-coupling range. The other systems exhibited sufficiently small $\lambda_{ep}$ values appropriate for weak-coupling superconductors, with the $\lambda_{ep}$ value of the Al–Zn–Mg i-QC being considerably smaller than those of the others. Furthermore, the materials with higher Ta concentrations tended to have higher $D(E_F)$ values, with the Al–Zn–Mg i-QC exhibiting the smallest value among the specimens.

Superconductivity in a quasiperiodic system has recently been investigated using an attractive Hubbard model on a Penrose lattice—a typical two-dimensional quasicrystalline lattice—using the real-space dynamical mean-field theory[17]. Unconventional spatially extended Cooper pairs were observed to form under weak-coupling conditions; the sum of the momenta of the Cooper pair electrons was nonzero, in contrast to the zero total momenta of the Cooper pair in conventional BCS superconductivity. Unlike any other known superconductors in periodic systems, the present finding opens up a new research field to investigate unprecedented unique superconducting states, spatially inhomogeneous on *any* length scale reflecting the self-similarity of the quasiperiodic structure[11].

In summary, polygrain single-phase $Ta_{1.6}Te$ dd-QC specimens were fabricated by reaction sintering and then subjected to electrical resistivity, magnetic susceptibility, and specific heat measurements in this study. The results unconditionally validated the occurrence of bulk superconductivity at a $T_c$ of ~1 K. This is the first example of superconductivity in thermodynamically stable QCs. The temperature dependence of the upper critical field $H_{c2}$ was consistent with the WHH theory in the dirty limit. The extrapolation of $H_{c2}$ to $T = 0$ yielded a value of 32 kOe. These findings are anticipated to motivate further investigations on the physical properties of vdW layered quasicrystals as well as two-dimensional quasicrystals, including unprecedented unique superconductivity theoretically expected for QCs.

**Methods**

**Sample synthesis**

$Ta_{1.6}Te$ samples were synthesized by reaction sintering. $TaTe_2$ and Ta were used as the starting materials at a ratio of 1:3. $TaTe_2$ was prepared in advance from a 1:2 elemental powder mixture of Ta (Rare Metallic; ~325 mesh, 99.9%) and Te (Rare Metallic; ~100 mesh, 99.99%). The mixture was pressed into a pellet of diameter 10 mm and inserted into a quartz tube, which was subsequently evacuated to $5-6 \times 10^{-4}$ Pa, sealed, and then subjected to two-step annealing (773 K for 24 h, followed by 1,273 K for 24 h). Subsequently, the mixture of the $TaTe_2$ powder, Ta powder, and a shot (30 – 40 mg) of iodine (FUJIFILM Wako Pure Chemical; 99.9%; used to promote the reaction) was pressed into a 5-mm-thick pellet of diameter 10 mm. The pellet was sealed in an evacuated ($5-6 \times 10^{-4}$ Pa) quartz tube and then sintered at 1,273 K for six days.

**Sample characterization**

The structures of the sintered samples were examined by powder XRD and electron diffractometry. Powder XRD profiles were obtained using a Rigaku diffractometer (RINT-2500V) with Cu $K\alpha$ radiation at 40 kV and 200 mA, a divergence slit open-angle of 0.5°, scattering slit open-angle of 0.5°, receiving slit width of 0.15 mm, rate of 0.5°/min, and step of 0.01°. Electron diffraction patterns were obtained using a TEM (JEM-2010F, JEOL) operated at an acceleration voltage of 200 kV.

**Physical property measurements**

The electrical resistivity was measured using a physical property measurement system (PPMS; Quantum Design) with a four-probe method and a direct current of 25 –100 μA.

Samples with cross-sectional areas of 2–3 mm × 2–3 mm and lengths of 5–6 mm were used for the electrical resistivity measurements. The magnetic susceptibility was measured using a magnetic property measurement system (MPMS3; Quantum Design) with a $^3$He refrigerator. The specific heat was measured using the PPMS with the thermal relaxation method.

## Acknowledgements

This study was supported by the JST-CREST program (grant no. JPMJCR22O3; Japan) and JSPS KAKENHI Grant Numbers JP19H05821 and JP23K04355). A part of this work was performed using facilities of the Cryogenic Research Center, the University of Tokyo. A part of this work was supported by "Advanced Research Infrastructure for Materials and Nanotechnology in Japan (ARIM)" of the Ministry of Education, Culture, Sports, Science and Technology (MEXT), Grant Number JPMXP1222UT0053.

## Author contributions

K.H., S.N., and Y.T. fabricated the samples and performed powder XRD measurements. Y.K. and K.E. analyzed the powder XRD data and constructed the profile. Y.K., K.H., and Y.T. performed the electron diffraction experiments. Y.T., K.H., and K.E. conducted the electrical resistivity and specific heat measurements. S.S., R.T., Y.T., and K.H. conducted the magnetic susceptibility experiments. K.E. designed and supervised the whole project. Y.T. and K.E. cowrote the manuscript. All the authors have discussed the results and contributed to the manuscript.

**Table 1** | Superconducting characteristics of the Ta$_{1.6}$Te dodecagonal quasicrystal (dd-QC), other Ta–chalcogenide systems, Ta, and the Al–Zn–Mg i-QC. $T_c$ and $\Theta_D$ were sourced from the literature. $\lambda_{ep}$ was calculated from $T_c$ and $\Theta_D$ using equation (2), where $\mu^*$ values of 0.1 and 0.13 were used for the Al–Zn–Mn i-QC and the other systems, respectively. $D(E_F)$ was calculated from $\lambda_{ep}$ and reported $\gamma$ using equation (3).

|  | $T_c$ (K) | $\Theta_D$ (K) | $\lambda_{ep}$ | $D(E_F)$ (10$^{47}$ states/J · m$^3$) | Reference |
|---|---|---|---|---|---|
| Ta$_{1.6}$Te dd-QC | 0.98 | 132 | 0.52 | 1.24 | Present study |
| Ta$_2$Se | 3.8 | 271 | 0.61 | 1.82 | [22] |
| TaSe$_2$ | 0.15 | 202 | 0.37 | 0.685 | [23] |
| TaS$_2$ | 0.8 | 236 | 0.45 | 1.19 | [23] |
| Ta | 4.4 | 231 | 0.67 | 2.39 | [24] |
| Al-Zn-Mg i-QC | 0.05 | 317 | 0.27 | 0.453 | [16] |

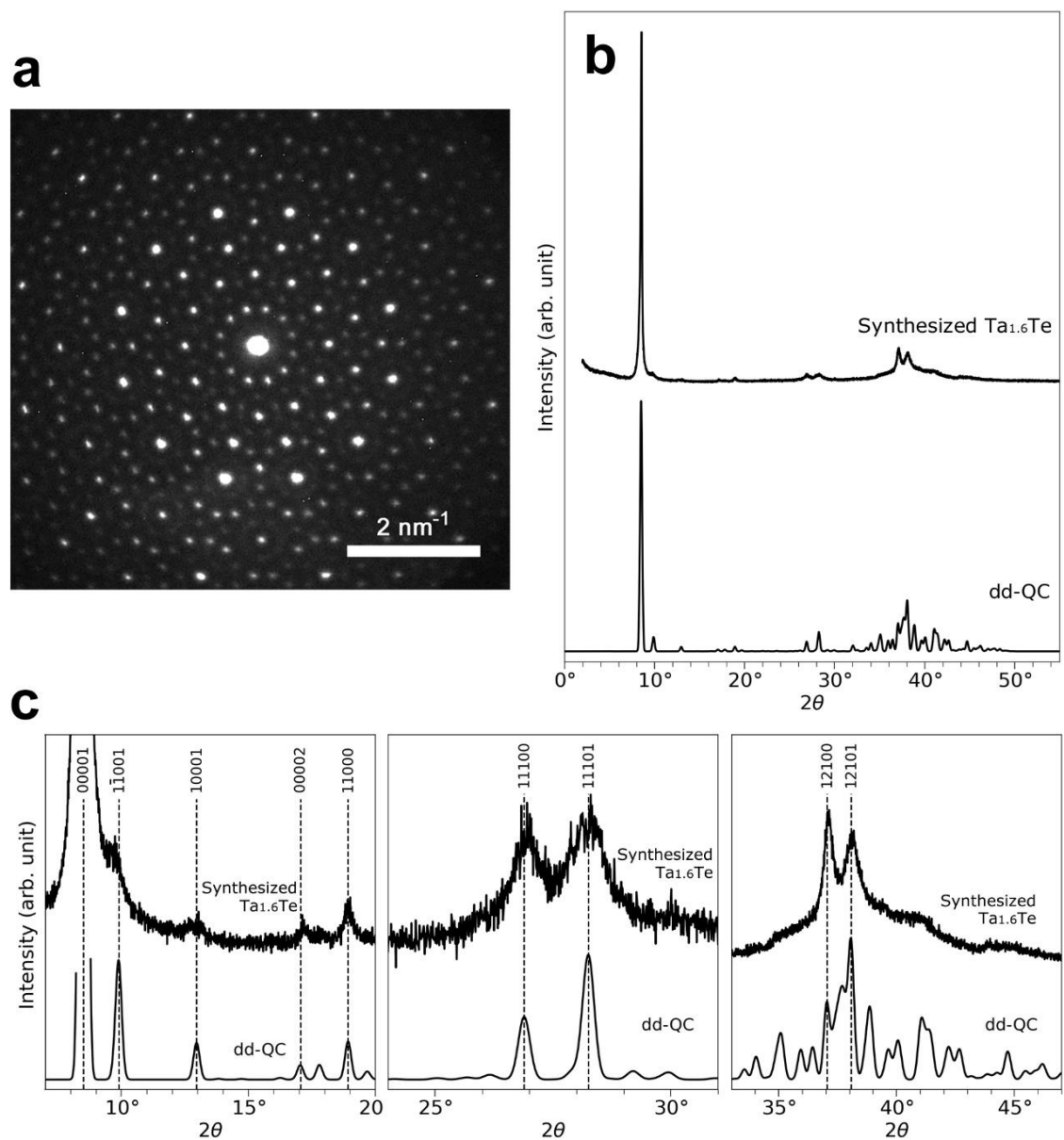

**Fig. 1 | Electron and X-ray diffraction data. a,** Electron diffraction pattern of the synthesized $Ta_{1.6}Te$. **b**, Experimentally obtained powder XRD profile of the synthesized $Ta_{1.6}Te$ and the calculated profile for the dd-QC phase. **c**, Magnified version of the data shown in **b**. The principles and procedures underlying the calculations are detailed in Supplementary Section 1.2.

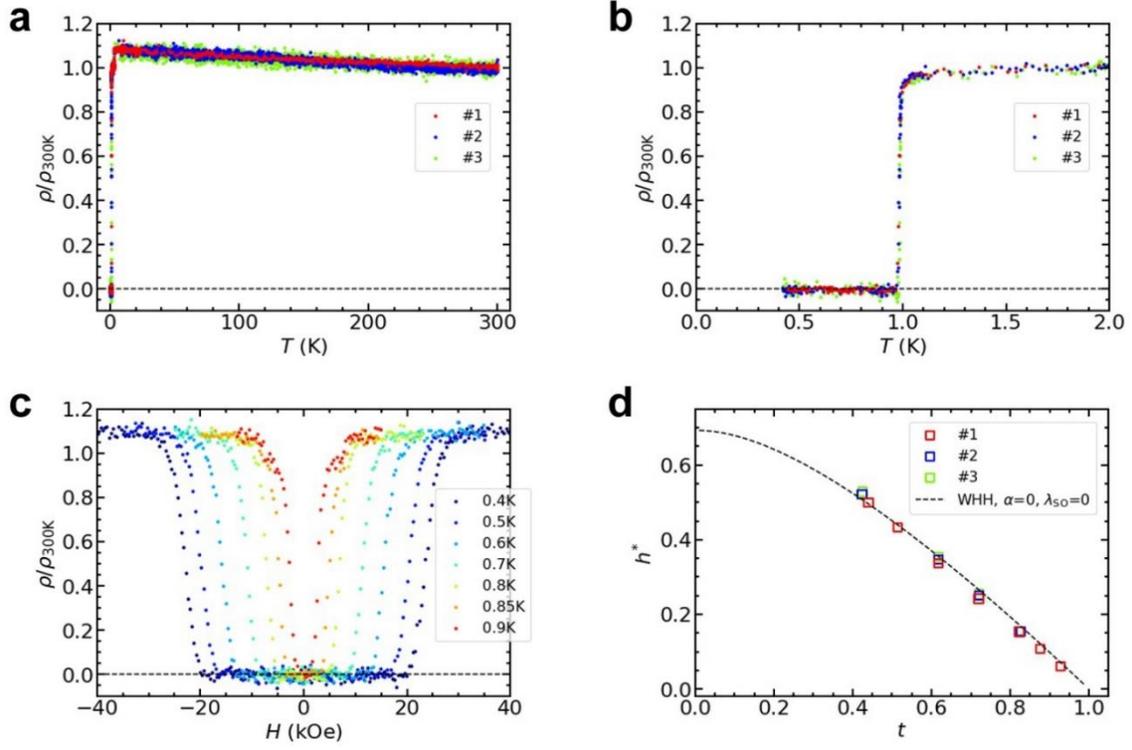

**Fig. 2 | Electrical resistivity measurements. a,** Temperature dependence of the electrical resistivity normalized by the value at $T = 300$ K—$\rho/\rho_{300\,K}$—for Samples #1–#3. **b,** Temperature dependence of $\rho/\rho_{300\,K}$ for Samples #1–#3 magnified at temperatures below 2 K. **c,** Magnetic field dependence of $\rho/\rho_{300\,K}$ for Sample #1 at different temperatures. **d,** Temperature dependence of the upper critical field $H_{c2}$ analyzed by plotting $h^* = -H_{c2}/\left[T_c(dH_{c2}/dT)_{T=T_c}\right]$ against $t = T/T_c$. Open squares represent the experimentally obtained data for Samples #1–#3, and the dashed curve represents the Werthamer–Helfand–Hohenberg (WHH) fit with $\alpha = 0$ and $\lambda_{SO} = 0$.

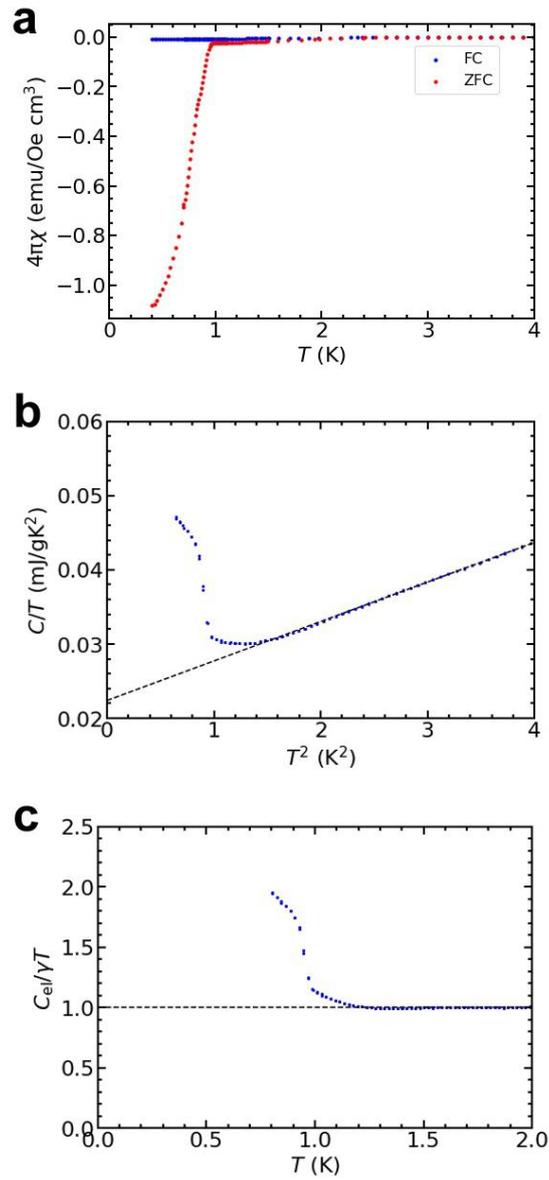

**Fig. 3 | Magnetic susceptibility and specific heat measurements.** Temperature dependence of **a,** the magnetic susceptibility $\chi$; **b,** the specific heat $C$, analyzed by plotting $C/T$ against $T^2$; and **c,** the ratio between the electronic component of the specific heat $C_{\text{el}}$ and $\gamma T$.

Supplementary information

# Superconductivity in a van der Waals layered quasicrystal

Yuki Tokumoto[1]*, Kotaro Hamano[1], Sunao Nakagawa[1], Yasushi Kamimura[1],

Shintaro Suzuki[2], Ryuji Tamura[3,4] & Keiichi Edagawa[1,4]*

[1]*Institute of Industrial Science, The University of Tokyo, Tokyo 153-8505, Japan*

[2]*Department of Physical Science, Aoyama Gakuin University, Kanagawa 252-5258,*

*Japan*

[3]*Department of Materials Science and Technology, Tokyo University of Science, Tokyo*

*125-8585, Japan*

[4]*CREST, Japan Science and Technology Agency (JST), Saitama 332-0012, Japan*

*e-mail: tokumoto@iis.u-tokyo.ac.jp; edagawa@iis.u-tokyo.ac.jp

**Supplementary Discussion 1. Analyses of the powder X-ray diffractometry profiles**

**1.1. XRD peak indexing**
The experimentally acquired X-ray diffractometry (XRD) profile ( Cu K$\alpha_1$; $\lambda$ = 1.5405 Å) was compared with those of the crystalline phases of the TaTe$_2$ and Ta materials employed in the reaction sintering (Supplementary Fig. 1). No traces of these phases were observed in the collected profile, indicating the completion of the reaction and formation of another phase.

      An attempt was made to index the diffraction peaks in the acquired profile for the dodecagonal quasicrystal (dd-QC). The electron diffraction pattern obtained from the sample revealed dodecagonal symmetry (Supplementary Fig. 2a). All the spots in the pattern were indexed using the four basis vectors $\mathbf{a}_i^*$ ($i$ = 1–4) shown in the figure, with $a^* = |\mathbf{a}_i^*| \approx 0.69$ Å$^{-1}$. The dd-QC phase has been reported to exhibit the periodic stacking of a two-dimensional (2D) quasicrystalline layer with a period of $c \approx 10.4$ Å (ref. 1). This yields $c^* = |\mathbf{c}^*| = \frac{2\pi}{c} \approx 0.604$ Å$^{-1}$, where the basis vector $\mathbf{c}^*$ is perpendicular to $\mathbf{a}_i^*$. Using the five basis vectors $\mathbf{a}_i^*$ ($i$ = 1–4) and $\mathbf{c}^*$, the peaks in the acquired XRD profile were successfully indexed, as shown in Supplementary Fig. 2b and Supplementary Table 1, with the values of $a^*$ and $c^*$ refined to 0.6942 and 0.6047 Å$^{-1}$, respectively.

**1.2. Construction of the XRD profiles**
In this subsection, the calculations performed to construct the profile of the dd-QC phase using one of its crystal approximant (CA) phases are detailed, and a comparison between the calculated and acquired profiles is presented.

*1.2.1. CA phases*
To date, the structures of two Ta–Te CA phases with compositions Ta$_{21}$Te$_{13}$ and Ta$_{97}$Te$_{60}$ have been determined using single-crystal XRD measurements[2,3]. The projections of their structures onto the *c*-plane (Supplementary Figs. 3a and b) show atomic clusters with dodecagonal symmetry that are arranged periodically. Between the CA phases Ta$_{21}$Te$_{13}$ and Ta$_{97}$Te$_{60}$, the latter was selected to calculate the dd-QC phase profile. This is an orthorhombic (pseudo-tetragonal) crystal with lattice parameters *a* = 27.672, *b* = 27.672, and *c* = 20.613 Å, with the unit cell in the *ab*-plane indicated by the blue square in Supplementary Fig. 3b. The basic structure of this phase is a tetragonal crystal with lattice parameters *a* = 19.567 and *c* = 10.307 Å, with the unit cell in the *ab*-plane indicated by the red square in Supplementary Fig. 3b. A weak superlattice order is introduced into the basic structure of the Ta$_{97}$Te$_{60}$ phase. The XRD profiles calculated using VESTA software[4] from the determined structures of the two CA phases[2,3] were compared with the acquired profile (Supplementary Fig. 4). The overall features of the collected profile resembled those calculated for the CA phases, highlighting the similarity between the dd-QC and CA phase structures. A more detailed comparison is presented in Supplementary Fig. 9.

*1.2.2. Principles underlying the calculation*
In general, by introducing phason strain to a QC structure, we can obtain a series of periodic structures, corresponding to CA structures to the QC one. Supplementary Fig. 5a shows a typical one-dimensional (1D) quasiperiodic structure, called the Fibonacci lattice, which is expressed as a 1D section of a 2D periodic structure. Here, $E_\parallel$ is the physical space, and $E_\perp$ is the complementary space perpendicular to $E_\parallel$. Furthermore, a line segment on the 2D square lattice spanned by $\mathbf{A}_1$ and $\mathbf{A}_2$ exhibits a periodic arrangement, and a 1D point sequence is formed on $E_\parallel$ as a set of line segment intersections with $E_\parallel$. The slope of $E_\parallel$ with respect to the 2D lattice is irrational ($\tau = (1+\sqrt{5})/2$ in this case), which leads to the point sequence on $E_\parallel$ being quasiperiodic instead of periodic.

A phason-strained Fibonacci lattice is shown in Supplementary Fig. 5b. A phason strain can be imposed onto the Fibonacci lattice by applying shear strain to the 2D structure, as indicated by the arrows on the left and right sides of Supplementary Fig. 5a. Upon introducing a shear strain, the vector $\mathbf{t} = \mathbf{A}_1 + 2\mathbf{A}_2$ shown in Supplementary Fig. 5a descends toward $E_\parallel$, as shown in Supplementary Fig. 5b. The structure resting atop $E_\parallel$ in Supplementary Fig. 5b exhibits periodicity with a period corresponding to its length. This periodic structure may be regarded as a 2/1 CA to the Fibonacci lattice, because the irrational slope of $\tau$ is replaced with a rational slope of 2/1.

The Fourier transform of the Fibonacci lattice can be calculated, as shown schematically in Supplementary Fig. 5c, where $E_\parallel^*$ and $E_\perp^*$ represent the physical reciprocal space and its complementary space, respectively. First, the Fourier transform of the 2D periodic structure shown in Supplementary Fig. 5a was calculated; this comprised $\delta$-functions at the 2D reciprocal lattice points ($\mathbf{G} = \mathbf{G}_\parallel + \mathbf{G}_\perp$; $\mathbf{G}_\parallel \in E_\parallel^*$, $\mathbf{G}_\perp \in E_\perp^*$) spanned by the basis vectors $\mathbf{A}_1^*$ and $\mathbf{A}_2^*$ shown in Supplementary Fig. 5c. Their intensities were determined through the Fourier transform of the line segment, which was in the form $f(\mathbf{G}_\perp) = a \sin(b|\mathbf{G}_\perp|)/(b|\mathbf{G}_\perp|)$ ($a, b$: constants) and independent of $\mathbf{G}_\parallel$. This indicates that the reciprocal lattice points with a large $|\mathbf{G}_\perp|$ always exhibited small intensities. The Fourier transform $F(\mathbf{q})$ of the Fibonacci lattice could be expressed as the projection of the $\delta$-functions at the 2D reciprocal lattice points onto $E_\parallel^*$, as shown at the bottom of Supplementary Fig. 5c, that is:

$$F(\mathbf{q}) = \sum_\mathbf{G} \delta(\mathbf{q} - \mathbf{G}_\parallel) \cdot f(\mathbf{G}_\perp). \quad (\text{S1})$$

Here, $\mathbf{G}_\parallel = h_1 \mathbf{a}_1^* + h_2 \mathbf{a}_2^*$ ($h_1, h_2 \in$ integers), where $\mathbf{a}_1^*$ and $\mathbf{a}_2^*$ are the projections of $\mathbf{A}_1^*$ and $\mathbf{A}_2^*$ onto $E_\parallel^*$, respectively. Because $|f(\mathbf{G}_\perp)|$ is always small for a large $|\mathbf{G}_\perp|$, a threshold value $G_\perp^0$ ($> 0$) could be used for $|\mathbf{G}_\perp|$, and the summation in equation (S1) for $\mathbf{G}$ could be limited to $|\mathbf{G}_\perp| < G_\perp^0$, that is:

$$F(\mathbf{q}) = \sum_{\mathbf{G} \text{ with } |\mathbf{G}_\perp|<G_\perp^0} \delta(\mathbf{q} - (h_1\mathbf{a}_1^* + h_2\mathbf{a}_2^*)) \cdot f(\mathbf{G}_\perp). \quad (\text{S2})$$

The Fourier transform of the 2/1 CA was performed based on the fundamental Fourier transformation attributes, as shown in Supplementary Fig. 5d. This corresponds to the projection of the Fourier transform of the 2D structure shown in Supplementary Fig. 5b. The 2D Fourier transform shown in Supplementary Fig. 5d was obtained by introducing a shear strain to the 2D Fourier transform shown in Supplementary Fig. 5c, as indicated by the arrows on the top and bottom of the figure, which led to the vector $\mathbf{t}^* = -2\mathbf{A}_1^* + \mathbf{A}_2^*$ descending toward $E_\perp^*$. Finally, the Fourier transform of the 2/1 CA was obtained by projecting the 2D Fourier transform onto $E_\parallel^*$, which comprised $\delta$-functions, as shown at the bottom of Supplementary Fig. 5d. If an appropriate value is adopted for $G_\perp^0$, no more than one 2D reciprocal lattice point will fall onto $E_\parallel^*$ at the same position.

Consequently, the Fourier transform of the 2/1 CA can be expressed in the same form as that of equation (S2), but with a different set of basis vectors.

$$F(\mathbf{q}) = \sum_{\mathbf{G} \text{ with } |\mathbf{G}_\perp| < G_\perp^0} \delta(\mathbf{q} - (h_1 \mathbf{b}_1^* + h_2 \mathbf{b}_2^*)) \cdot f(\mathbf{G}_\perp), \quad (S3)$$

where $\mathbf{b}_1^*$ and $\mathbf{b}_2^*$ are the projections of the basis vectors $\mathbf{B}_1^*$ and $\mathbf{B}_2^*$, respectively, in the 2D reciprocal lattice shown in Supplementary Fig. 5d. Here, the position of the $\delta$-function, $h_1 \mathbf{b}_1^* + h_2 \mathbf{b}_2^*$, can be reindexed using only one basis vector $\mathbf{d}_1^* = \mathbf{b}_1^*$ as $H_1 \mathbf{d}_1^*$, given the relationship $H_1 = h_1 + 2h_2$. Appropriate selection of $G_\perp^0$ assures one-to-one correspondence between $(h_1, h_2)$ and $H_1$. Consequently, if the diffraction intensity function $(I(\mathbf{q}) \equiv |F(\mathbf{q})|^2)$ for the Fibonacci lattice can be determined, that for the 2/1 CA can be obtained simply by shifting each $\delta$-function from $h_1 \mathbf{a}_1^* + h_2 \mathbf{a}_2^*$ to $H_1 \mathbf{d}_1^* = (h_1 + 2h_2)\mathbf{d}_1^*$ without changing the intensity and vice versa. This argument should generally hold true for a QC and its CA. Subsequently, this scheme was applied to the dd-QC phase and one of its CA phases, and the XRD profile of the former was calculated using that of the latter.

### 1.2.3. Four-dimensional description of a dd-QC and its CAs[5]

For a dd-QC and its CAs, a four-dimensional (4D) space spanned by orthonormal unit vectors $\mathbf{e}_i$ ($i = 1 - 4$) can be considered, with $\mathbf{e}_1$ and $\mathbf{e}_2$ spanning the physical space $E_\parallel$ ($E_\parallel^*$) and $\mathbf{e}_3$ and $\mathbf{e}_4$ encompassing the complementary space $E_\perp$ ($E_\perp^*$); these can be compared with those in Supplementary Figs. 5a–d. A 4D dd lattice is spanned by $\mathbf{A}_i$ ($i = 1 - 4$) expressed as:

$$\mathbf{A}_i = \sum_{j=1}^{4} M_{ij} \mathbf{e}_j,$$

$$M = \frac{a_{4D}}{\sqrt{2}} \begin{pmatrix} \sqrt{3}/2 & -1/2 & -\sqrt{3}/2 & -1/2 \\ 1 & 0 & 1 & 0 \\ 0 & 1 & 0 & 1 \\ -1/2 & \sqrt{3}/2 & -1/2 & -\sqrt{3}/2 \end{pmatrix}. \quad (S4)$$

Here, $a_{4D}$ is the lattice constant of the 4D lattice. The lattice reciprocal to this 4D lattice is spanned by $\mathbf{A}_i^*$ ($i = 1 - 4$) expressed as:

$$\mathbf{A}_i^* = \sum_{j=1}^{4} {}^t M_{ij}^{-1} \mathbf{e}_j,$$

$${}^t M_{ij}^{-1} = \frac{a_{4D}^*}{\sqrt{2}} \begin{pmatrix} 1 & 0 & -1 & 0 \\ \sqrt{3}/2 & 1/2 & \sqrt{3}/2 & -1/2 \\ 1/2 & \sqrt{3}/2 & -1/2 & \sqrt{3}/2 \\ 0 & 1 & 0 & -1 \end{pmatrix}. \quad (S5)$$

Here, the superscript $t$ denotes transposition, and $a_{4D}^*$ ($= 4\pi/(\sqrt{3} a_{4D})$) represents the lattice constant of the 4D reciprocal lattice. Supplementary Fig. 6 shows the projections of $\mathbf{A}_i$ ($\mathbf{A}_i^*$) ($i = 1 - 4$)—$\mathbf{a}_i$ ($\mathbf{a}_i^*$) and $\mathbf{u}_i$ ($\mathbf{u}_i^*$)—onto $E_\parallel$ ($E_\parallel^*$) and $E_\perp$ ($E_\perp^*$), respectively.

A series of square CAs can be obtained by imposing a phason strain onto the 4D dodecagonal lattice, which causes the following two vectors $\mathbf{t}_1$ and $\mathbf{t}_2$ to descend toward $E_\parallel$ (see Supplementary Figs. 5a and b):

$$\mathbf{t}_1 = (2p, q, p, 0)_{\mathbf{A}_i}$$
$$\mathbf{t}_2 = (0, p, q, 2p)_{\mathbf{A}_i}, \quad (S6)$$

where $p$ and $q$ are integers, and the subscript $\mathbf{A}_i$ indicates that the vectors are represented with respect to the basis vectors $\mathbf{A}_i$ ($i = 1 - 4$). The vectors $\mathbf{t}_1$ and $\mathbf{t}_2$ can be rewritten using equation (S4) with the basis vectors $\mathbf{e}_i$ ($i = 1 - 4$) as:

$$\mathbf{t}_1 = \frac{a_{4D}}{\sqrt{2}}(\sqrt{3}p + q, 0, -\sqrt{3}p + q, 0)_{\mathbf{e}_i}$$
$$\mathbf{t}_2 = \frac{a_{4D}}{\sqrt{2}}(0, \sqrt{3}p + q, 0, -\sqrt{3}p + q)_{\mathbf{e}_i}, \quad (S7)$$

Equation (S7) indicates that (1) the $E_\perp$ components of $\mathbf{t}_1$ and $\mathbf{t}_2$ approach zero as $q/p \rightarrow \sqrt{3}$; (2) the $E_\parallel$ components of $\mathbf{t}_1$ and $\mathbf{t}_2$ are oriented along $\mathbf{e}_1$ and $\mathbf{e}_2$, respectively, corresponding to the lattice translational vectors of the square CA structure; and (3) the lattice constant of the square lattice can be expressed as:

$$a = \frac{\sqrt{3}p + q}{\sqrt{2}} a_{4D}. \quad (S8)$$

Point (1) suggests that the closer $q/p$ is to $\sqrt{3}$, the more identical is the CA structure to the QC structure.

The introduction of the phason strain changes the basis vectors of the 4D dodecagonal lattice $\mathbf{A}_i$ ($i = 1 - 4$) to $\mathbf{B}_i$ ($i = 1 - 4$). Correspondingly, the basis vectors of the reciprocal lattice $\mathbf{A}_i^*$ ($i = 1 - 4$) are transformed into $\mathbf{B}_i^*$ ($i = 1 - 4$), similar to the case of the Fibonacci lattice shown in Supplementary Figs. 5a–d. On the other hand, equation (S7) and the relationship $a_{4D}^* = 4\pi/(\sqrt{3}a_{4D})$ indicate that the reciprocal basis vectors of the square lattice of the CA—$\mathbf{d}_1^*$ and $\mathbf{d}_2^*$—can be expressed with respect to the basis vectors $\mathbf{e}_1$ and $\mathbf{e}_2$ as:

$$\mathbf{d}_1^* = \frac{a_{4D}^*}{\sqrt{2}}\left(\frac{1}{p + q/\sqrt{3}}, 0\right)$$
$$\mathbf{d}_2^* = \frac{a_{4D}^*}{\sqrt{2}}\left(0, \frac{1}{p + q/\sqrt{3}}\right), \quad (S9)$$

Based on the relationship between the projections of $\mathbf{B}_i^*$ ($i = 1 - 4$) onto $E_\parallel^*$ (that is, $\mathbf{b}_i^*$ ($i = 1 - 4$)) and $\mathbf{d}_i^*$ ($i = 1,2$), the $\delta$-functions at $h_1\mathbf{a}_1^* + h_2\mathbf{a}_2^* + h_3\mathbf{a}_3^* + h_4\mathbf{a}_4^*$ for the dd-QC are linked to those at $H_1\mathbf{d}_1^* + H_2\mathbf{d}_2^*$ for the CA through the expression:

$$\begin{pmatrix} H_1 \\ H_2 \end{pmatrix} = \begin{pmatrix} 2p & q & p & 0 \\ 0 & p & q & 2p \end{pmatrix} \begin{pmatrix} h_1 \\ h_2 \\ h_3 \\ h_4 \end{pmatrix}. \quad (S10)$$

### *1.2.4. Parameter determination*
$a^*$ was determined to be 0.6942 Å$^{-1}$ for the investigated dd-QC phase (see Section 1.1). This value should be equal to $\frac{a_{4D}^*}{\sqrt{2}}$ in equation (S5), which leads to $a_{4D}^* = 0.9817$ Å$^{-1}$. Furthermore, the relationship $a_{4D}^* = 4\pi/(\sqrt{3}a_{4D})$ yields an $a_{4D}$ value of 7.390 Å. The lattice constant ($a$) of the basic structure of the Ta$_{97}$Te$_{60}$ CA phase was determined to be 19.567 Å (Section 1.2.1)[3]. Notably, for $p = 1$ and $q = 2$, equation (S8) yields $a = 19.50$ Å, which is consistent with the lattice constant of the CA phase. Therefore, the Ta$_{97}$Te$_{60}$ phase was confirmed to be a square CA phase with $p = 1$ and $q = 2$ (tetragonal in 3D). Considering the superlattice ordering in the *ab*-plane and along the *c*-direction in the CA phase, as described in Section 1.2.1, the indices $(H_1', H_2', H_3')$ for the superlattice-ordered structure are related to those $(H_1, H_2, H_3)$ for the basic structure as follows:

$$\begin{pmatrix} H_1' \\ H_2' \\ H_3' \end{pmatrix} = \begin{pmatrix} 1 & -1 & 0 \\ 1 & 1 & 0 \\ 0 & 0 & 2 \end{pmatrix} \begin{pmatrix} H_1 \\ H_2 \\ H_3 \end{pmatrix}. \quad (S11)$$

Using equations (S11) and (S10) with $p = 1$ and $q = 2$, the relationship between the indices $(H_1', H_2', H_3')$ for the superlattice-ordered Ta$_{97}$Te$_{60}$ CA phase and those $(h_1, h_2, h_3, h_4, h_5)$ for the dd-QC phase can be obtained.

$$\begin{pmatrix} H_1' \\ H_2' \\ H_3' \end{pmatrix} = \begin{pmatrix} 2 & 1 & -1 & -2 & 0 \\ 2 & 3 & 3 & 2 & 0 \\ 0 & 0 & 0 & 0 & 2 \end{pmatrix} \begin{pmatrix} h_1 \\ h_2 \\ h_3 \\ h_4 \\ h_5 \end{pmatrix} \quad (S12)$$

$G_\perp^0$ was assumed to be $2.4 \frac{a_{4D}^*}{\sqrt{2}}$, and the XRD profile of the dd-QC phase was calculated using that of the Ta$_{97}$Te$_{60}$ CA phase with equation (S12).

### *1.2.5. Results*
Supplementary Table 2 presents the indices, $q$-value, the structural factor squared $|F|^2$, scattering angle $2\theta$, and powder diffraction intensity $I$ calculated for the Ta$_{97}$Te$_{60}$ CA phase using VESTA software[4] with the structural data reported by Harbrecht and Conrad[3], where $q = (4\pi \sin \theta)/\lambda$, and $\lambda = 1.5405$ Å$^{-1}$ (CuK$\alpha_1$). Supplementary Table 3 summarizes the calculation results of the dd-QC phase. Supplementary Figs. 7a and 7b show the diffraction patterns calculated for the Ta$_{97}$Te$_{60}$ CA and dd-QC phases on the planes perpendicular to the *c*-axis, respectively. Supplementary Fig. 8 shows the collected powder XRD profile of the sample and those calculated for the dd-QC, Ta$_{97}$Te$_{60}$ CA, and Ta$_{21}$Te$_{13}$ CA phases; Supplementary Figs. 9a–d present the magnified versions of these profiles.

**Supplementary Table 1 | XRD peak indexing results.** $q_{exp}$ and $q_{calc}$ denote the experimental and calculated $q$-values, respectively. $q = \frac{4\pi \sin\theta}{\lambda}$, and $\lambda = 1.5405$ Å.

| $q_{exp}$ | Index | $q_{calc}$ |
|---|---|---|
| 0.608 | 00001 | 0.6047 |
| 0.688 | 1$\bar{1}$001 | 0.7034 |
| 0.912 | 10001 | 0.9206 |
| 1.220 | 00002 | 1.2094 |
| 1.340 | 11000 | 1.3411 |
| 1.901 | 11100 | 1.8966 |
| 1.991 | 11101 | 1.9906 |
| 2.596 | 12100 | 2.5908 |
| 2.665 | 12101 | 2.6604 |

**Supplementary Table 2 | Powder XRD data for the Ta$_{97}$Te$_{60}$ crystal approximant (CA) phase.** Indices, $q$-value, scattering angle $2\theta$, and relative intensity $I$ calculated for the Ta$_{97}$Te$_{60}$ CA phase using VESTA[4] software with the structural data reported by Harbrecht and Conrad[3]. $q = \frac{4\pi \sin\theta}{\lambda}$ and $\lambda = 1.5405$ Å.

| H1' | H2' | H3' | q | 2θ | I |
|---|---|---|---|---|---|
| 0 | 0 | 2 | 0.6096 | 8.57 | 100.00 |
| 1 | 0 | 2 | 0.6505 | 9.15 | 11.93 |
| 1 | 1 | 2 | 0.6890 | 9.69 | 2.36 |
| 3 | 1 | 0 | 0.7180 | 10.10 | 0.36 |
| 2 | 0 | 2 | 0.7602 | 10.69 | 0.74 |
| 2 | 1 | 2 | 0.7934 | 11.16 | 2.05 |
| 2 | 2 | 2 | 0.8855 | 12.46 | 0.74 |
| 3 | 1 | 2 | 0.9419 | 13.26 | 1.03 |
| 0 | 0 | 4 | 1.2193 | 17.19 | 0.49 |
| 5 | 2 | 0 | 1.2228 | 17.24 | 0.23 |
| 1 | 0 | 4 | 1.2402 | 17.49 | 0.26 |
| 1 | 1 | 4 | 1.2608 | 17.78 | 0.27 |
| 5 | 3 | 0 | 1.3240 | 18.68 | 1.10 |
| 6 | 0 | 0 | 1.3624 | 19.23 | 0.71 |
| 3 | 1 | 4 | 1.4150 | 19.98 | 0.25 |
| 7 | 2 | 2 | 1.7619 | 24.95 | 0.22 |
| 8 | 2 | 0 | 1.8724 | 26.54 | 2.64 |
| 8 | 2 | 1 | 1.8970 | 26.90 | 0.95 |
| 6 | 6 | 0 | 1.9267 | 27.32 | 1.23 |
| 8 | 1 | 2 | 1.9295 | 27.37 | 0.42 |
| 6 | 6 | 1 | 1.9506 | 27.67 | 0.48 |
| 8 | 2 | 2 | 1.9691 | 27.94 | 5.13 |
| 7 | 1 | 4 | 2.0160 | 28.62 | 0.31 |
| 6 | 6 | 2 | 2.0208 | 28.69 | 2.57 |
| 8 | 3 | 2 | 2.0335 | 28.87 | 0.23 |
| 7 | 2 | 4 | 2.0540 | 29.17 | 0.32 |
| 8 | 2 | 3 | 2.0838 | 29.60 | 0.59 |
| 6 | 6 | 3 | 2.1327 | 30.31 | 0.31 |
| 6 | 5 | 4 | 2.1521 | 30.60 | 0.37 |
| 8 | 1 | 4 | 2.1995 | 31.29 | 0.24 |
| 8 | 2 | 4 | 2.2344 | 31.80 | 1.33 |
| 5 | 3 | 6 | 2.2578 | 32.14 | 0.35 |
| 6 | 6 | 4 | 2.2801 | 32.46 | 0.54 |
| 6 | 0 | 6 | 2.2805 | 32.47 | 0.22 |
| 8 | 3 | 4 | 2.2913 | 32.63 | 0.61 |
| 7 | 5 | 4 | 2.3026 | 32.79 | 0.21 |
| 6 | 2 | 6 | 2.3253 | 33.13 | 0.54 |
| 8 | 4 | 4 | 2.3688 | 33.76 | 0.43 |
| 9 | 1 | 4 | 2.3904 | 34.08 | 0.81 |
| 8 | 2 | 5 | 2.4143 | 34.43 | 0.68 |
| 9 | 2 | 4 | 2.4226 | 34.55 | 1.49 |
| 7 | 1 | 6 | 2.4337 | 34.72 | 1.95 |
| 0 | 0 | 8 | 2.4385 | 34.79 | 0.24 |
| 10 | 3 | 2 | 2.4477 | 34.92 | 0.42 |
| 6 | 4 | 6 | 2.4547 | 35.03 | 1.98 |
| 6 | 6 | 5 | 2.4566 | 35.06 | 0.34 |
| 1 | 1 | 8 | 2.4596 | 35.10 | 0.82 |
| 8 | 5 | 4 | 2.4648 | 35.18 | 0.27 |
| 9 | 3 | 4 | 2.4752 | 35.33 | 1.23 |
| 7 | 5 | 5 | 2.4775 | 35.36 | 0.72 |
| 2 | 0 | 8 | 2.4805 | 35.41 | 1.07 |
| 8 | 7 | 2 | 2.4895 | 35.54 | 0.38 |
| 2 | 1 | 8 | 2.4908 | 35.56 | 0.24 |
| 9 | 5 | 3 | 2.5102 | 35.85 | 0.22 |
| 5 | 3 | 7 | 2.5111 | 35.86 | 0.21 |
| 7 | 3 | 6 | 2.5170 | 35.95 | 2.91 |
| 2 | 2 | 8 | 2.5217 | 36.02 | 1.41 |
| 8 | 4 | 5 | 2.5392 | 36.27 | 0.70 |
| 10 | 3 | 3 | 2.5408 | 36.30 | 0.44 |
| 3 | 1 | 8 | 2.5420 | 36.32 | 2.87 |
| 9 | 4 | 4 | 2.5471 | 36.39 | 0.39 |
| 11 | 2 | 1 | 2.5568 | 36.54 | 0.23 |
| 7 | 7 | 4 | 2.5572 | 36.54 | 0.81 |
| 9 | 1 | 5 | 2.5594 | 36.57 | 0.88 |
| 8 | 8 | 0 | 2.5689 | 36.71 | 3.23 |
| 6 | 2 | 7 | 2.5720 | 36.76 | 0.30 |
| 10 | 0 | 4 | 2.5772 | 36.84 | 1.87 |
| 8 | 0 | 6 | 2.5777 | 36.84 | 1.10 |
| 8 | 7 | 3 | 2.5811 | 36.89 | 0.57 |
| 8 | 8 | 1 | 2.5869 | 36.98 | 4.50 |
| 10 | 1 | 4 | 2.5872 | 36.99 | 0.65 |
| 8 | 1 | 6 | 2.5877 | 36.99 | 0.20 |
| 11 | 3 | 0 | 2.5889 | 37.01 | 7.54 |
| 4 | 0 | 8 | 2.6022 | 37.21 | 2.02 |
| 11 | 3 | 1 | 2.6068 | 37.27 | 8.44 |
| 11 | 2 | 2 | 2.6108 | 37.33 | 0.80 |
| 10 | 4 | 3 | 2.6109 | 37.34 | 0.42 |
| 10 | 2 | 4 | 2.6170 | 37.43 | 1.82 |
| 8 | 2 | 6 | 2.6174 | 37.43 | 3.07 |
| 3 | 3 | 8 | 2.6219 | 37.50 | 1.35 |
| 9 | 5 | 4 | 2.6366 | 37.72 | 1.67 |
| 9 | 3 | 5 | 2.6387 | 37.75 | 0.66 |
| 8 | 8 | 2 | 2.6402 | 37.77 | 6.32 |
| 4 | 2 | 8 | 2.6415 | 37.79 | 2.40 |
| 6 | 6 | 6 | 2.6565 | 38.01 | 1.73 |
| 11 | 3 | 2 | 2.6597 | 38.06 | 13.34 |
| 10 | 3 | 4 | 2.6657 | 38.15 | 1.48 |
| 7 | 1 | 7 | 2.6703 | 38.22 | 0.60 |
| 11 | 4 | 1 | 2.6751 | 38.29 | 0.23 |
| 7 | 5 | 6 | 2.6758 | 38.30 | 3.47 |
| 6 | 4 | 7 | 2.6895 | 38.50 | 0.56 |
| 5 | 0 | 8 | 2.6899 | 38.51 | 0.62 |
| 11 | 2 | 3 | 2.6983 | 38.63 | 0.70 |
| 5 | 1 | 8 | 2.6994 | 38.65 | 2.58 |
| 8 | 7 | 4 | 2.7042 | 38.72 | 1.28 |
| 7 | 7 | 5 | 2.7158 | 38.89 | 0.50 |
| 11 | 4 | 2 | 2.7267 | 39.06 | 0.79 |
| 8 | 8 | 3 | 2.7268 | 39.06 | 3.16 |
| 5 | 2 | 8 | 2.7279 | 39.08 | 0.30 |
| 10 | 4 | 4 | 2.7326 | 39.15 | 1.64 |
| 8 | 4 | 6 | 2.7330 | 39.15 | 3.41 |
| 10 | 0 | 5 | 2.7347 | 39.18 | 1.15 |
| 11 | 3 | 3 | 2.7456 | 39.34 | 6.66 |
| 7 | 3 | 7 | 2.7464 | 39.35 | 0.65 |
| 9 | 1 | 6 | 2.7518 | 39.43 | 2.67 |
| 4 | 4 | 8 | 2.7561 | 39.50 | 0.50 |
| 10 | 2 | 5 | 2.7721 | 39.74 | 0.92 |
| 5 | 3 | 8 | 2.7748 | 39.78 | 2.87 |
| 11 | 1 | 4 | 2.7886 | 39.98 | 1.29 |
| 9 | 5 | 5 | 2.7907 | 40.01 | 0.80 |
| 6 | 0 | 8 | 2.7933 | 40.05 | 1.10 |
| 12 | 1 | 2 | 2.8013 | 40.17 | 0.56 |
| 10 | 6 | 3 | 2.8014 | 40.17 | 0.25 |
| 8 | 0 | 7 | 2.8022 | 40.19 | 0.23 |
| 6 | 1 | 8 | 2.8025 | 40.19 | 0.55 |
| 11 | 4 | 3 | 2.8106 | 40.31 | 0.37 |
| 11 | 2 | 4 | 2.8162 | 40.40 | 0.78 |
| 10 | 3 | 5 | 2.8182 | 40.43 | 0.44 |
| 9 | 3 | 6 | 2.8258 | 40.54 | 2.53 |
| 6 | 2 | 8 | 2.8300 | 40.60 | 1.90 |
| 3 | 1 | 9 | 2.8358 | 40.69 | 0.31 |
| 8 | 2 | 7 | 2.8388 | 40.73 | 0.51 |
| 5 | 4 | 8 | 2.8391 | 40.74 | 0.40 |
| 8 | 8 | 4 | 2.8435 | 40.80 | 2.64 |
| 8 | 7 | 5 | 2.8546 | 40.97 | 0.24 |
| 11 | 3 | 4 | 2.8616 | 41.08 | 6.11 |
| 12 | 3 | 2 | 2.8740 | 41.26 | 0.53 |
| 6 | 3 | 8 | 2.8751 | 41.28 | 0.38 |
| 10 | 4 | 5 | 2.8815 | 41.37 | 0.90 |
| 12 | 1 | 3 | 2.8830 | 41.40 | 0.31 |
| 7 | 5 | 7 | 2.8927 | 41.54 | 0.65 |
| 7 | 7 | 6 | 2.8978 | 41.62 | 0.73 |
| 12 | 2 | 3 | 2.9097 | 41.80 | 0.39 |
| 10 | 6 | 4 | 2.9152 | 41.88 | 0.45 |
| 10 | 0 | 6 | 2.9156 | 41.89 | 1.95 |
| 7 | 1 | 8 | 2.9196 | 41.95 | 1.44 |
| 11 | 4 | 4 | 2.9240 | 42.01 | 0.40 |
| 11 | 1 | 5 | 2.9347 | 42.17 | 0.34 |
| 12 | 4 | 2 | 2.9361 | 42.19 | 0.23 |
| 6 | 4 | 8 | 2.9372 | 42.21 | 1.50 |
| 8 | 4 | 7 | 2.9457 | 42.34 | 0.39 |
| 7 | 2 | 8 | 2.9460 | 42.34 | 0.27 |
| 10 | 2 | 6 | 2.9507 | 42.42 | 1.29 |
| 11 | 2 | 5 | 2.9610 | 42.57 | 0.23 |
| 9 | 1 | 7 | 2.9632 | 42.60 | 0.29 |
| 9 | 5 | 6 | 2.9681 | 42.68 | 0.67 |
| 8 | 8 | 5 | 2.9870 | 42.96 | 0.66 |
| 7 | 3 | 8 | 2.9894 | 43.00 | 1.26 |
| 11 | 5 | 4 | 3.0023 | 43.19 | 0.33 |
| 11 | 3 | 5 | 3.0042 | 43.22 | 1.54 |
| 12 | 2 | 4 | 3.0194 | 43.45 | 0.30 |
| 9 | 3 | 7 | 3.0320 | 43.64 | 0.24 |
| 8 | 0 | 8 | 3.0407 | 43.77 | 0.32 |
| 5 | 3 | 9 | 3.0461 | 43.86 | 0.28 |
| 0 | 0 | 10 | 3.0482 | 43.89 | 0.30 |
| 10 | 4 | 6 | 3.0537 | 43.97 | 0.28 |
| 1 | 0 | 10 | 3.0566 | 44.02 | 0.81 |
| 1 | 1 | 10 | 3.0650 | 44.14 | 0.23 |
| 8 | 2 | 8 | 3.0745 | 44.29 | 0.62 |
| 2 | 0 | 10 | 3.0818 | 44.40 | 0.26 |
| 2 | 1 | 10 | 3.0902 | 44.52 | 0.29 |
| 13 | 1 | 3 | 3.0985 | 44.65 | 0.24 |
| 11 | 1 | 6 | 3.1040 | 44.73 | 0.21 |
| 6 | 6 | 8 | 3.1078 | 44.79 | 0.30 |
| 2 | 2 | 10 | 3.1151 | 44.90 | 0.79 |
| 7 | 5 | 8 | 3.1244 | 45.04 | 0.67 |
| 3 | 1 | 10 | 3.1316 | 45.15 | 2.28 |
| 11 | 3 | 6 | 3.1697 | 45.73 | 0.25 |
| 8 | 4 | 8 | 3.1735 | 45.79 | 0.25 |
| 4 | 0 | 10 | 3.1806 | 45.90 | 0.79 |
| 9 | 1 | 8 | 3.1897 | 46.04 | 0.20 |
| 3 | 3 | 10 | 3.1968 | 46.15 | 0.53 |
| 14 | 2 | 0 | 3.2111 | 46.37 | 0.41 |
| 4 | 2 | 10 | 3.2129 | 46.39 | 0.24 |
| 14 | 2 | 1 | 3.2255 | 46.59 | 0.37 |
| 5 | 0 | 10 | 3.2527 | 47.00 | 0.33 |
| 9 | 3 | 8 | 3.2537 | 47.02 | 0.27 |
| 5 | 1 | 10 | 3.2606 | 47.12 | 0.77 |
| 14 | 2 | 2 | 3.2685 | 47.24 | 0.43 |
| 14 | 4 | 0 | 3.3060 | 47.82 | 0.41 |
| 14 | 4 | 1 | 3.3201 | 48.04 | 0.40 |
| 5 | 3 | 10 | 3.3233 | 48.09 | 0.46 |
| 6 | 0 | 10 | 3.3388 | 48.32 | 0.35 |
| 14 | 4 | 2 | 3.3618 | 48.68 | 0.57 |
| 14 | 4 | 3 | 3.4302 | 49.74 | 0.25 |

**Supplementary Table 3 | Calculated powder XRD data for the dodecagonal quasicrystal (dd-QC) phase.** Indices, $q$-value, scattering angle $2\theta$, and relative intensity $I$ calculated for the dd-QC phase. $q = \frac{4\pi \sin\theta}{\lambda}$ and $\lambda = 1.5405$ Å.

| h1 | h2 | h3 | h4 | h5 | q | 2θ | I |
|---|---|---|---|---|---|---|---|
| 0 | 0 | 0 | 0 | 1 | 0.6047 | 8.50 | 100.00 |
| 1 | 0 | 0 | 0 | 0 | 0.6942 | 9.76 | 0.48 |
| 1 | -1 | 0 | 0 | 1 | 0.7034 | 9.89 | 5.47 |
| 1 | 0 | 0 | 0 | 1 | 0.9206 | 12.96 | 1.77 |
| 0 | 0 | 0 | 0 | 2 | 1.2094 | 17.05 | 0.49 |
| 1 | -1 | 0 | 0 | 2 | 1.2617 | 17.80 | 0.71 |
| 1 | 1 | 0 | 0 | 0 | 1.3411 | 18.93 | 1.81 |
| 1 | 0 | 0 | 0 | 2 | 1.3945 | 19.69 | 0.36 |
| 1 | 1 | 1 | 0 | 0 | 1.8966 | 26.89 | 3.87 |
| 2 | 0 | 0 | 1 | 2 | 1.9678 | 27.92 | 0.52 |
| 1 | 1 | 1 | 0 | 1 | 1.9906 | 28.25 | 7.70 |
| 2 | 1 | -1 | 0 | 2 | 2.0555 | 29.19 | 0.48 |
| 2 | 1 | 0 | 0 | 1 | 2.1082 | 29.96 | 0.42 |
| 2 | 1 | -1 | -1 | 2 | 2.2494 | 32.02 | 1.87 |
| 1 | 1 | 0 | 0 | 3 | 2.2560 | 32.11 | 0.56 |
| 2 | 0 | 0 | 0 | 3 | 2.2845 | 32.53 | 0.55 |
| 2 | 1 | 0 | 0 | 2 | 2.3541 | 33.55 | 1.44 |
| 2 | 0 | 0 | 1 | 3 | 2.3876 | 34.04 | 3.27 |
| 0 | 0 | 0 | 0 | 4 | 2.4189 | 34.50 | 0.24 |
| 1 | -1 | 0 | 0 | 4 | 2.4454 | 34.89 | 2.70 |
| 2 | 1 | -1 | 0 | 3 | 2.4604 | 35.11 | 6.05 |
| 1 | 0 | 0 | 0 | 4 | 2.5165 | 35.94 | 4.28 |
| 2 | 2 | -1 | 0 | 2 | 2.5506 | 36.44 | 4.67 |
| 1 | 2 | 1 | 0 | 0 | 2.5908 | 37.04 | 10.76 |
| 1 | 0 | 0 | 1 | 4 | 2.6105 | 37.33 | 4.43 |
| 2 | 2 | -1 | -1 | 2 | 2.6188 | 37.45 | 2.26 |
| 2 | 1 | -1 | -1 | 3 | 2.6245 | 37.54 | 4.81 |
| 2 | -1 | -1 | 0 | 4 | 2.6351 | 37.70 | 7.69 |
| 3 | 1 | -1 | 0 | 2 | 2.6433 | 37.82 | 4.73 |
| 1 | 2 | 1 | 0 | 1 | 2.6604 | 38.07 | 19.60 |
| 2 | 0 | -1 | 0 | 4 | 2.7012 | 38.68 | 3.08 |
| 2 | 1 | 0 | 0 | 3 | 2.7148 | 38.88 | 9.54 |
| 2 | 2 | 0 | 0 | 1 | 2.7495 | 39.40 | 0.25 |
| 1 | 1 | 0 | 0 | 4 | 2.7658 | 39.64 | 3.97 |
| 3 | 1 | 0 | 0 | 1 | 2.7729 | 39.75 | 0.40 |
| 2 | 0 | 0 | 0 | 4 | 2.7890 | 39.99 | 2.40 |
| 3 | 1 | -1 | -1 | 2 | 2.7968 | 40.10 | 3.80 |
| 2 | 2 | 0 | -1 | 2 | 2.8592 | 41.04 | 7.89 |
| 2 | 0 | 0 | 1 | 4 | 2.8741 | 41.26 | 3.84 |
| 2 | 2 | -1 | 0 | 3 | 2.8868 | 41.45 | 5.29 |
| 3 | 1 | -1 | -2 | 1 | 2.9195 | 41.94 | 0.48 |
| 2 | 1 | -1 | 0 | 4 | 2.9348 | 42.18 | 3.70 |
| 2 | 2 | 0 | 0 | 2 | 2.9422 | 42.29 | 0.59 |
| 2 | 2 | -1 | -1 | 3 | 2.9473 | 42.36 | 1.20 |
| 3 | 1 | 0 | 0 | 2 | 2.9641 | 42.62 | 1.49 |
| 3 | 1 | -1 | 0 | 3 | 2.9691 | 42.69 | 2.99 |
| 3 | 2 | -1 | 0 | 1 | 3.0009 | 43.17 | 0.43 |
| 0 | 0 | 0 | 0 | 5 | 3.0236 | 43.52 | 0.20 |
| 1 | -1 | 0 | 0 | 5 | 3.0449 | 43.84 | 0.72 |
| 3 | 2 | -1 | -1 | 1 | 3.0591 | 44.05 | 0.23 |
| 2 | 1 | -1 | -1 | 4 | 3.0737 | 44.28 | 0.92 |
| 3 | 1 | -1 | -2 | 2 | 3.1017 | 44.70 | 0.45 |
| 1 | 0 | 0 | 0 | 5 | 3.1023 | 44.71 | 3.07 |
| 3 | 1 | -1 | -1 | 3 | 3.1065 | 44.77 | 0.57 |
| 2 | 1 | 0 | 0 | 4 | 3.1512 | 45.45 | 1.12 |
| 2 | 2 | 0 | -1 | 3 | 3.1628 | 45.63 | 0.25 |
| 1 | 0 | 0 | 1 | 5 | 3.1790 | 45.88 | 0.94 |
| 2 | -1 | -1 | 0 | 5 | 3.1992 | 46.18 | 2.07 |
| 3 | 2 | -1 | -2 | 0 | 3.2108 | 46.36 | 0.28 |
| 2 | 0 | -1 | 0 | 5 | 3.2539 | 47.02 | 0.78 |
| 3 | 2 | -1 | -2 | 1 | 3.2672 | 47.22 | 0.21 |
| 3 | 2 | 0 | -1 | 0 | 3.2849 | 47.50 | 0.71 |
| 2 | 2 | -1 | 0 | 4 | 3.3005 | 47.74 | 0.61 |
| 1 | 1 | 0 | 0 | 5 | 3.3077 | 47.85 | 0.49 |
| 3 | 2 | 0 | -1 | 1 | 3.3401 | 48.35 | 0.90 |

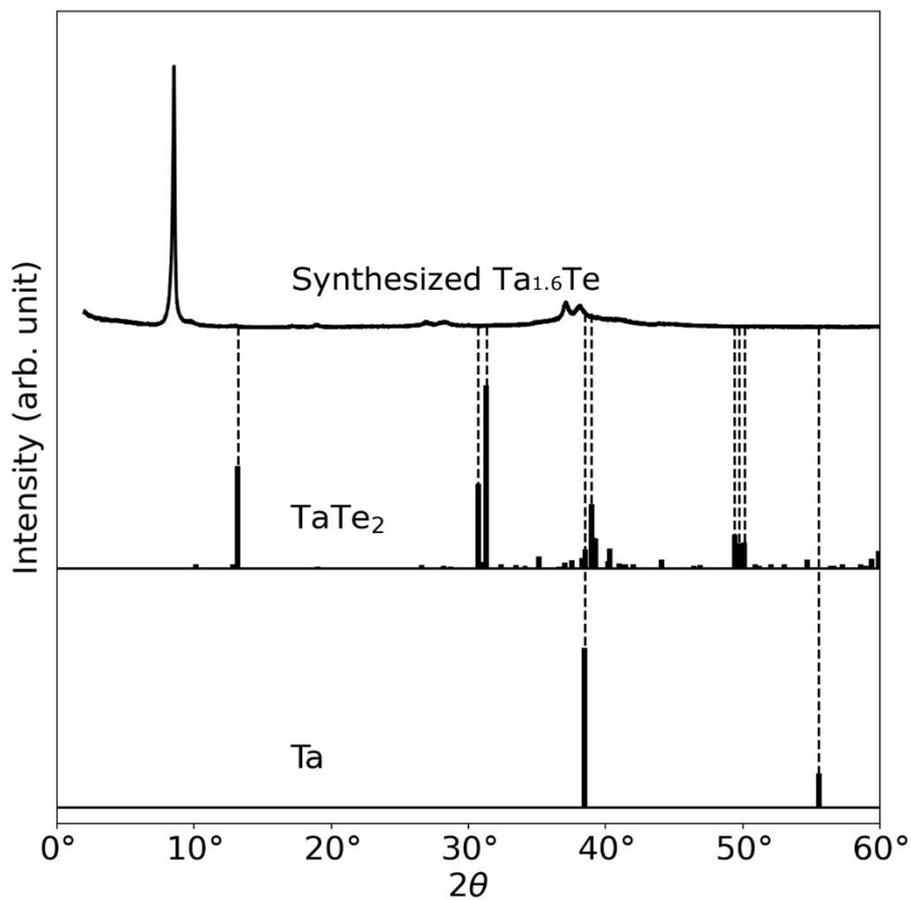

**Supplementary Fig. 1 | Powder XRD analysis.** Comparison between the experimentally acquired powder XRD profile of the $Ta_{1.6}Te$ sample (Cu $K\alpha_1$; $\lambda = 1.5405$ Å) and the peak data of the crystalline phases of $TaTe_2$ and Ta obtained from 'Powder Diffraction Datafile (PDF)'.

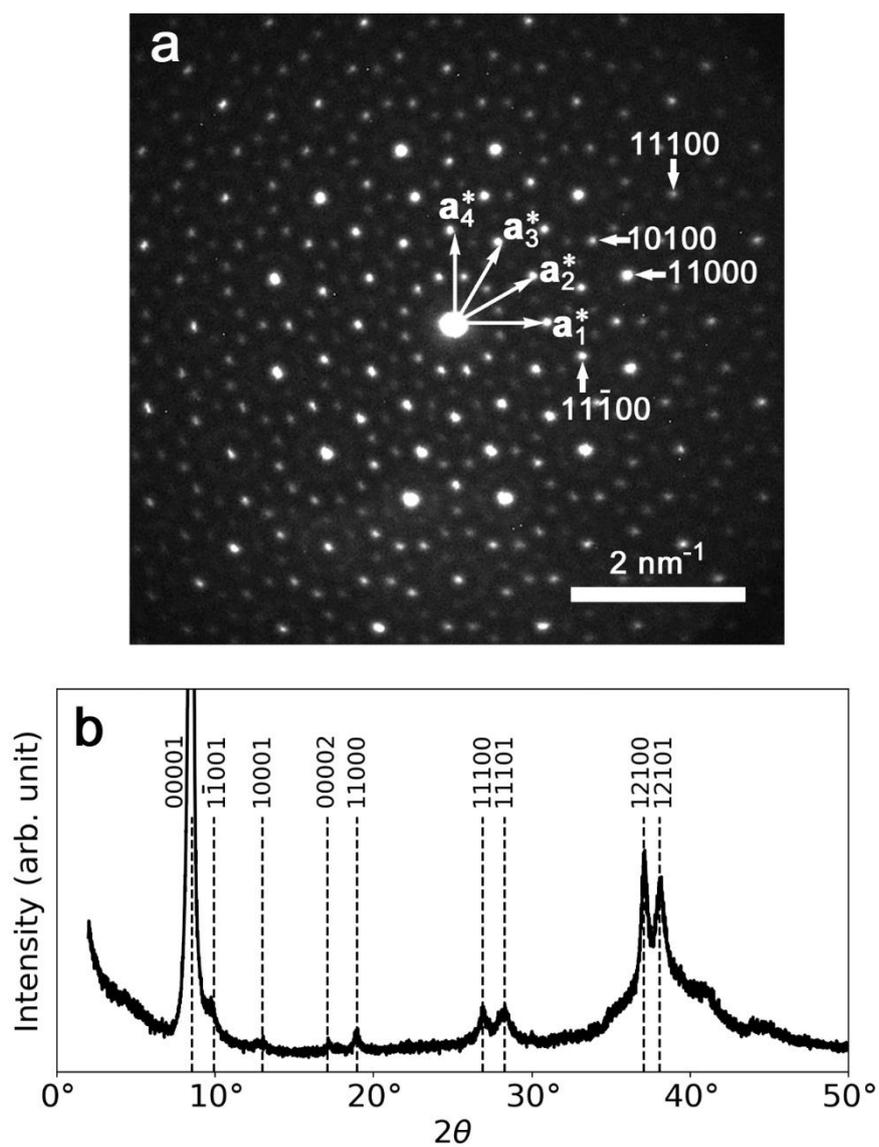

**Supplementary Fig. 2 | XRD peak indexing. a**, A twelve-fold electron diffraction pattern showing the four basis vectors $\mathbf{a}_i^*$ ($i$ = 1–4) used for indexing. **b**, Results of the XRD peak indexing.

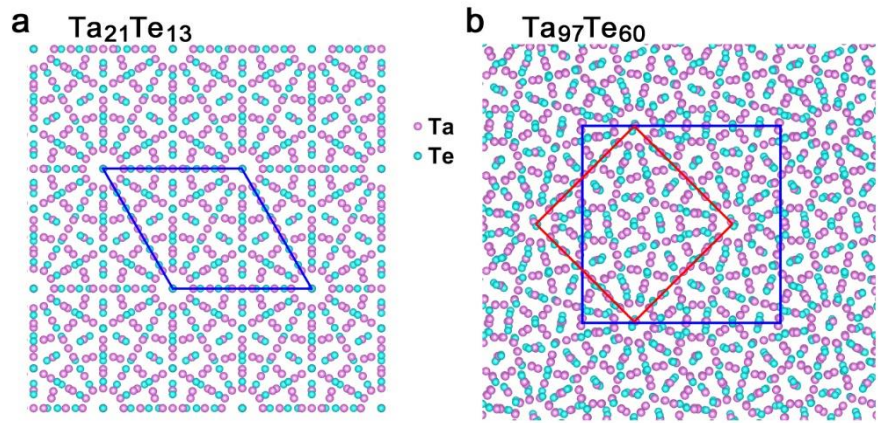

**Supplementary Fig. 3 | Structures of the CA phases. a, b**, Projections of the $Ta_{21}Te_{13}$ (**a**) and $Ta_{97}Te_{60}$ (**b**) CA phase structures onto the *c*-plane[2,3]. The unit cells of the crystals are indicated in blue, and that of the basic structure in **b** is shown in red.

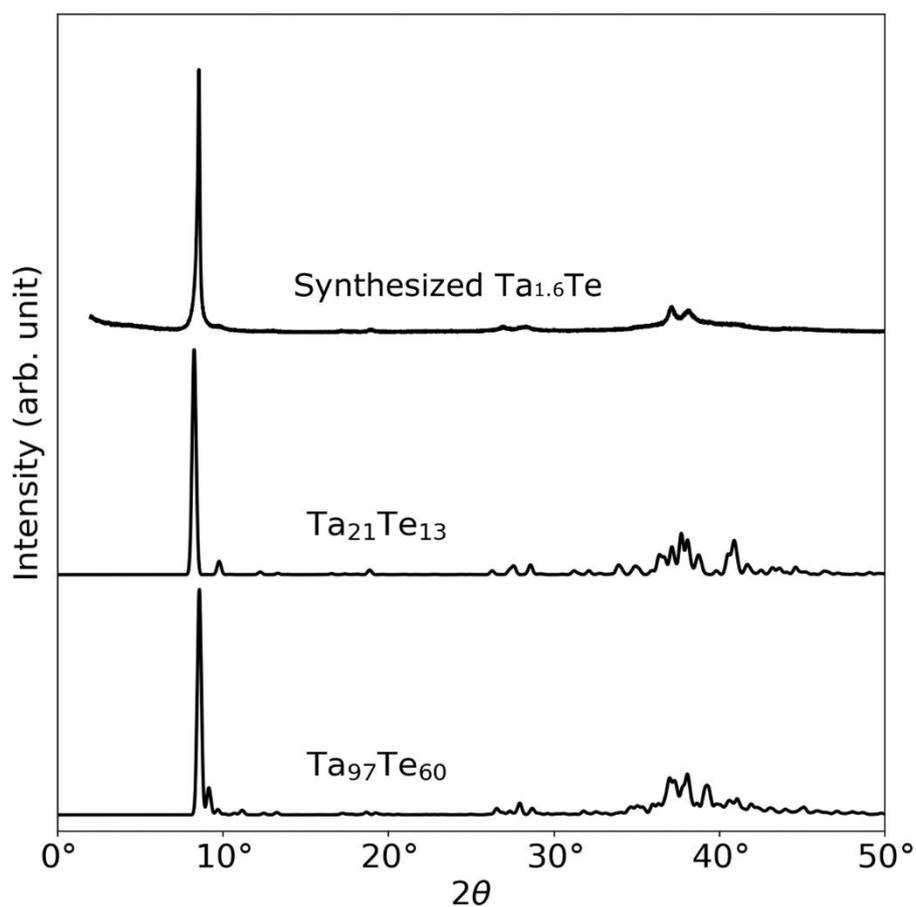

**Supplementary Fig. 4 | Powder XRD investigation.** Experimentally obtained powder XRD profile of the $Ta_{1.6}Te$ sample (Cu $K\alpha_1$; $\lambda = 1.5405$ Å) and calculated profiles of the $Ta_{21}Te_{13}$ and $Ta_{97}Te_{60}$ CA phases. Gaussians with a full width at half maximum of $\Delta q = 0.02$ Å$^{-1}$ were used for the calculated profiles.

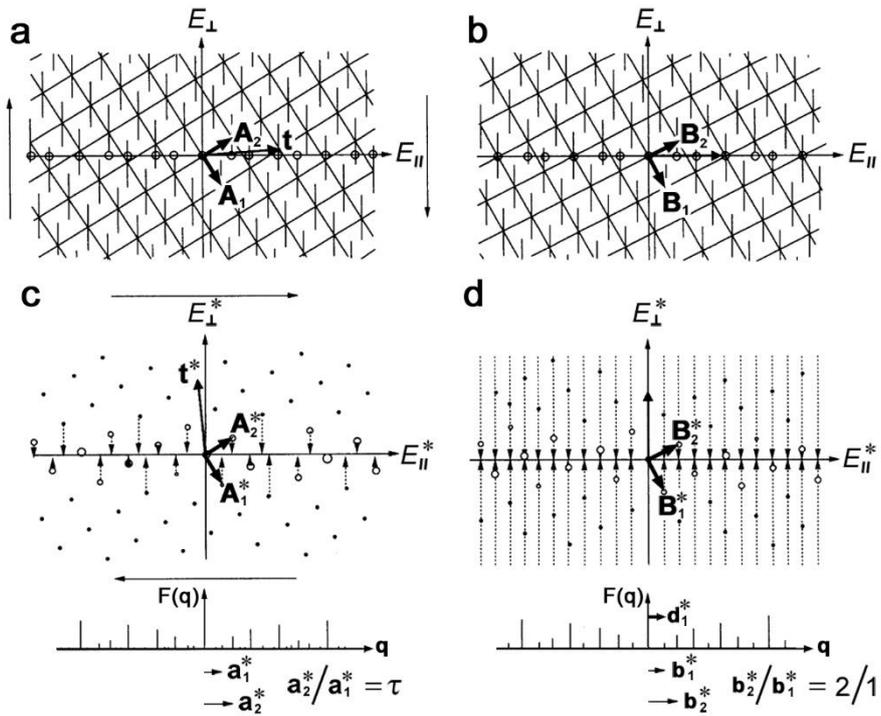

**Supplementary Fig. 5 | Relationships between a QC and its CA in real and reciprocal spaces a**, A 1D quasiperiodic structure, called a Fibonacci lattice, expressed as a 1D section of a 2D periodic structure. **b**, A CA obtained upon imposing a phason strain onto the Fibonacci lattice shown in (**a**). **c**, **d**, Fourier transforms of the **c**, Fibonacci lattice and **d**, CA.

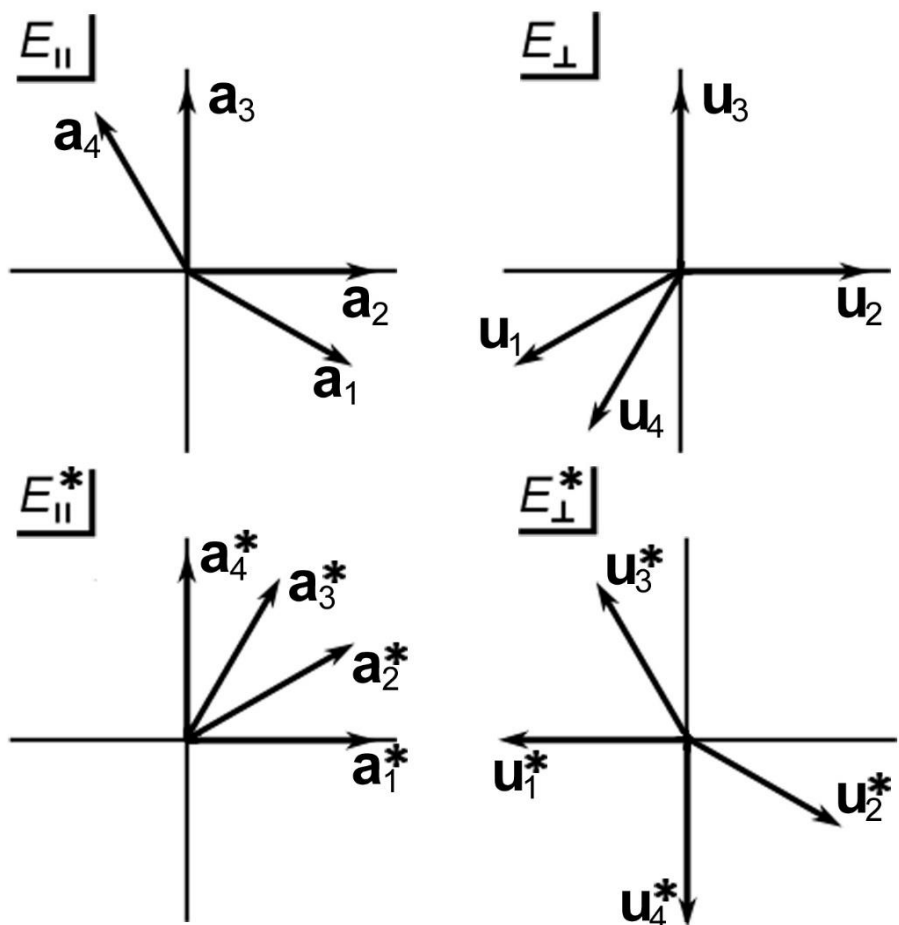

**Supplementary Fig. 6 | Projections of the basis vectors of a 4D dodecagonal lattice.** Projections of the basis vectors $\mathbf{A}_i$ ($\mathbf{A}_i^*$) ($i = 1 - 4$) of a 4D dodecagonal lattice (that is, $\mathbf{a}_i$ ($\mathbf{a}_i^*$) and $\mathbf{u}_i$ ($\mathbf{u}_i^*$)) onto $E_{||}$ ($E_{||}^*$) and $E_\perp$ ($E_\perp^*$), respectively.

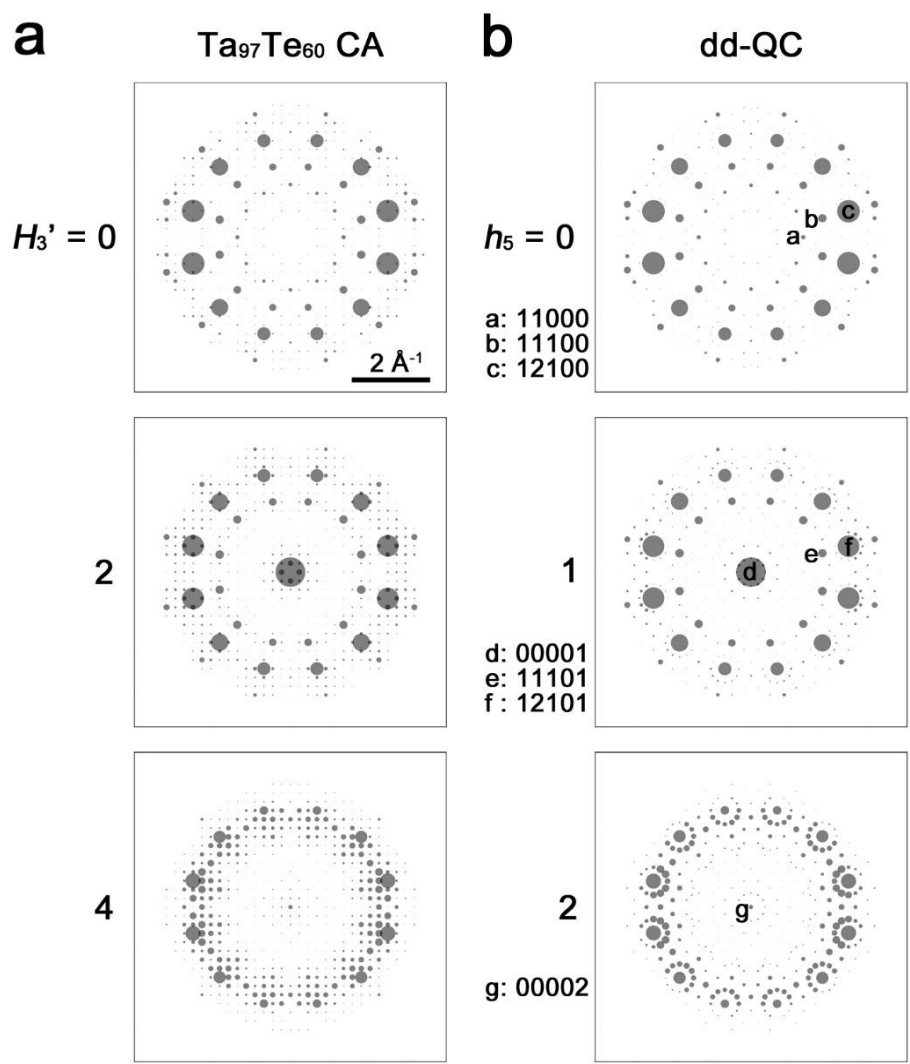

**Supplementary Fig. 7 | Diffraction patterns of the CA and dd-QC phases. a, b**, Calculated diffraction patterns of the **a**, $Ta_{97}Te_{60}$ CA and **b**, dd-QC phases on the planes perpendicular to the *c*-axis. The area of the circle is proportional to $|F|^2$.

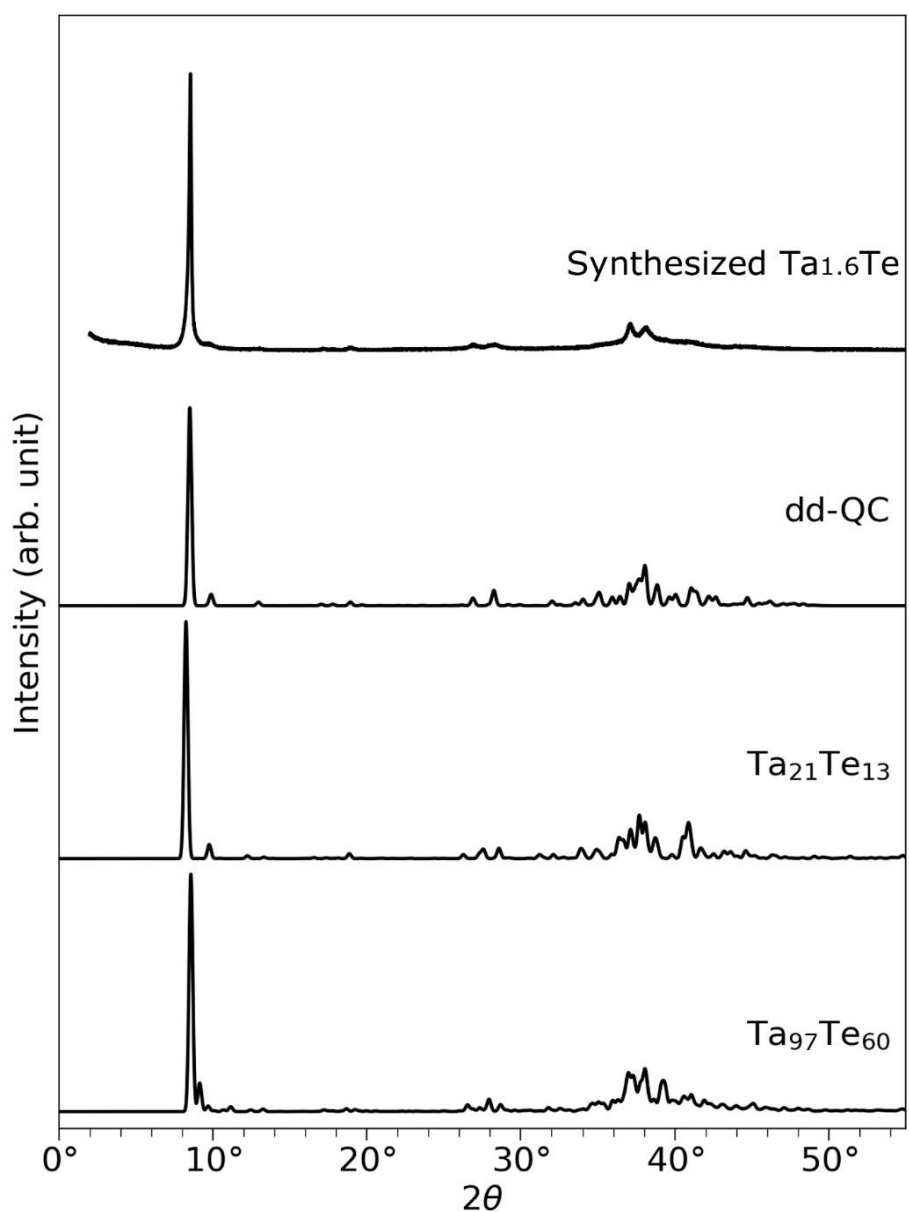

**Supplementary Fig. 8 | Powder XRD analysis.** Powder XRD profiles acquired for the synthesized $Ta_{1.6}Te$ sample and calculated for the dd-QC, $Ta_{21}Te_{13}$ CA, and $Ta_{97}Te_{60}$ CA phases. Gaussians with a full width at half maximum of $\Delta q = 0.02$ Å$^{-1}$ were used for the calculated profiles.

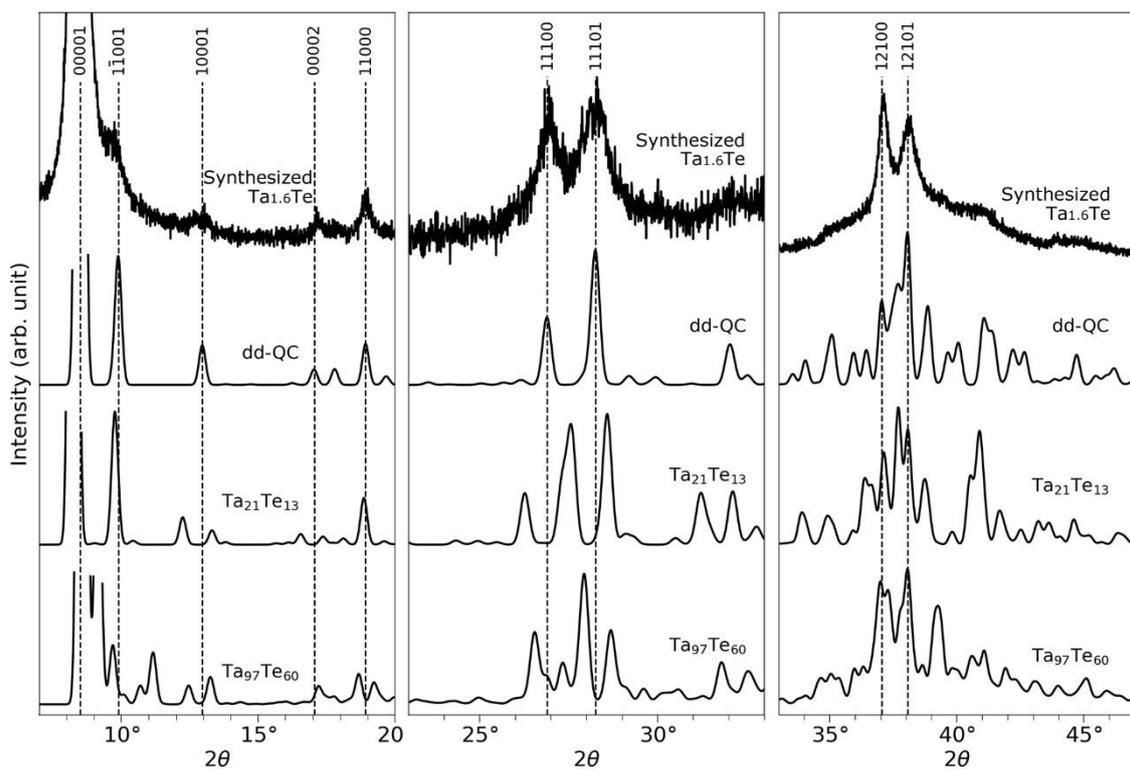

**Supplementary Fig. 9 | Magnified versions of the powder XRD profiles shown in Supplementary Fig. 8.** Powder XRD profiles acquired for the synthesized $Ta_{1.6}Te$ sample and calculated for the dd-QC, $Ta_{21}Te_{13}$ CA, and $Ta_{97}Te_{60}$ CA phases. Gaussians with a full width at half maximum of $\Delta q = 0.02\ \text{Å}^{-1}$ were used for the calculated profiles.